\documentclass[prl,aps,twocolumn,showpacs,superscriptaddress,footnoteinbib]{revtex4-1}

\usepackage{graphicx}
\usepackage{dcolumn}
\usepackage{bm}
\usepackage{amssymb,amsmath}
\usepackage{sidecap}
\usepackage{wrapfig}
\usepackage[bookmarks]{hyperref}
\usepackage{natbib}

\usepackage{graphicx}
\usepackage{leftidx}
\usepackage{color}

\usepackage{bbold} 

\usepackage{wasysym} 

\newcommand{\be}{\begin{equation}}
\newcommand{\ee}{\end{equation}}
\newcommand{\ba}{\begin{align}}
\newcommand{\ea}{\end{align}}
\newcommand{\sysb}{\left\{\begin{array}}
\newcommand{\syse}{\end{array}\right.}
\newcommand{\baa}{\begin{array}}
\newcommand{\eaa}{\end{array}}

\newcommand{\matb}{\left(\begin{array}}
\newcommand{\mate}{\end{array}\right)}

\newcommand{\ran}{\right\rangle}
\newcommand{\ket}[1]{\left| #1 \ran}

\definecolor{myblue}{rgb}{0,0,0.75}

\newcommand{\Hs}{H}
\newcommand{\rs}{\rho}
\newcommand{\ts}{\tau}

\begin{document}

\title{Robustness of many-body localization in the presence of dissipation}

\author{Emanuele Levi}
\affiliation{School of Physics and Astronomy, University of Nottingham, Nottingham, NG7 2RD, UK}
\author{Markus Heyl}
\affiliation{Physik Department, Technische Universit\"at M\"unchen, 85747 Garching, Germany}
\author{Igor Lesanovsky}
\author{Juan P. Garrahan}
\affiliation{School of Physics and Astronomy, University of Nottingham, Nottingham, NG7 2RD, UK}

\begin{abstract}
Many-body localization (MBL) has emerged as a novel paradigm for robust ergodicity breaking in closed quantum many-body systems. However, it is not yet clear to which extent MBL survives in the presence of dissipative processes induced by the coupling to an environment. Here we study heating and ergodicity for a paradigmatic MBL system---an interacting fermionic chain subject to quenched disorder---in the presence of dephasing. We find that, even though the system is eventually driven into an infinite-temperature state, heating as monitored by the von Neumann entropy can progress logarithmically slowly, implying exponentially large time scales for relaxation.  This slow loss of memory of initial conditions make signatures of non-ergodicity visible over a long, but transient, time regime. We point out a potential controlled realization of the considered setup with cold atomic gases held in optical lattices.
\end{abstract}


\maketitle
\textit{Introduction.---}
Within statistical physics thermodynamic systems relax to thermal states which are independent of their initial conditions \cite{Peliti2011,Eisert2014,Dalessio2015}. 
There exist, however, generic many-body systems violating this paradigm, for example classical glasses \cite{Binder2011,Biroli2013}. Recently, many-body localization (MBL) has emerged as a novel prototype for robust \emph{quantum} ergodicity breaking~\cite{Altshuler1997,Basko2006,*Gornyi2005,*Oganesyan2007,Pal2010}
which has attracted a lot of interest \cite{[See e.g.\ ] Canovi2011,*Bardarson2012,*Serbyn2013,*Huse2014,[See e.g.\ ]Imbrie2014,*Scardicchio2015,*Andraschko2014,*Yao2014,*Serbyn2014,*Vasseur2015,*Agarwal2015,*Laumann2014,[See e.g.\ ]De-Roeck2014,*Yao2014b,*Horssen2015,
[For a review see ]Altman2015,[For a review see ]Nandkishore2015}. While experiments have indeed demonstrated ergodicity breaking compatible with MBL~\cite{Schreiber2015,Bordia2015,Smith2015}, a significant challenge remains: it is not clear whether the imperfect isolation from the environment will eventually induce ergodicity on long time scales and therefore  destroy the MBL state.

Here we address the associated open question: what survives of MBL in the presence of dissipation? For that purpose we study the dynamics in a paradigmatic MBL system \cite{Nandkishore2015}, a chain of interacting fermions subject to disorder, in the presence of a Markovian particle-number preserving bath. We find that dissipation leads in the long-time limit to infinite heating and therefore destroys the MBL phase. Most importantly, however, the heating dynamics itself is extremely slow. In particular, the system's entropy increases only logarithmically in time, implying exponentially large relaxation time scales.  

Moreover, we show that this slow heating is reflected in the dynamics of the spin imbalance that has already been measured experimentally  in related systems~\cite{Schreiber2015,Bordia2015}. This means that there is a large time window where the non-ergodic character of MBL becomes apparent, before ultimate relaxation to the trivial infinite temperature state.  A further signature of this slow relaxing regime is a pattern of emissions into the bath that is intermittent both in space and in time.  We provide an outlook on how the reported phenomena can be observed experimentally with cold atoms. In this context, we show that the diagonal entropy~\cite{Polkovnikov2011} exhibits the same qualitative properties as the full entropy with the advantage that it is much easier to access experimentally.


\textit{Model.---}
We study the influence of dissipation for a paradigmatic MBL system, an open chain of interacting fermions in a random potential, with Hamiltonian
\begin{equation}
 \Hs=-J\sum_{l=1}^{N}\left(c^\dagger_l c_{l+1} + c^\dagger_{l+1} c_{l}\right) + V\sum_{l=1}^N n_l n_{l+1}+2\sum_{l=1}^Nh_l n_l,
 \label{eq:defHamiltonian}
\end{equation}
where $c^\dagger_l$ creates a fermion on site $l=1,\dots,N$ with $N$ the number of lattice sites, $n_l=c^\dagger_l c_l$ is the local number operator, and the local random potentials $h_l \in \left[-h,h\right]$ are drawn from uncorrelated uniform distributions. This model exhibits many-body localization transition at infinite temperature~\cite{Pal2010,Luitz2015,Serbyn2015b}, and potentially a many-body mobility edge for decreasing energy density~\cite{Luitz2015,Serbyn2015b} (whose existence has been questioned~\cite{Roeck2015}).

We study ergodicity of the MBL system in Eq.~(\ref{eq:defHamiltonian}) in the presence of a Markovian particle-number preserving bath which can be interpreted as a structureless environment allowing for energy exchange at all scales. Specifically, we consider a scenario where the full dynamics of the system can be described within a quantum Master equation of Lindblad form \cite{Lindblad1976,Gardiner2004}
\begin{equation}
 \dot{\rs}(t)=-i\left[\Hs,\rs(t)\right]+\gamma\sum_{l=1}^N\left[n_{l}\rs(t)n_l-\frac{1}{2}\{n_{l},\rs(t)\}\right],
 \label{eq:defMasterEquation}
\end{equation}
where $\rs$ is the system's density matrix and  $\gamma\geq0$ sets the coupling to the bath. As initial state we choose a charge-density wave type state $|\psi_0\rangle = |1010 \dots 10\rangle$, $\rs(t=0) = | \psi_0 \rangle \langle \psi_0 |$, with every second lattice site occupied, which is of particular experimental importance~\cite{Schreiber2015,Bordia2015,Smith2015}. Below we discuss the experimental relevance of this model system.

\begin{figure*}[ht]
\centering
\includegraphics[trim = 0mm 0mm 0mm 0mm, clip, width=1.75\columnwidth]{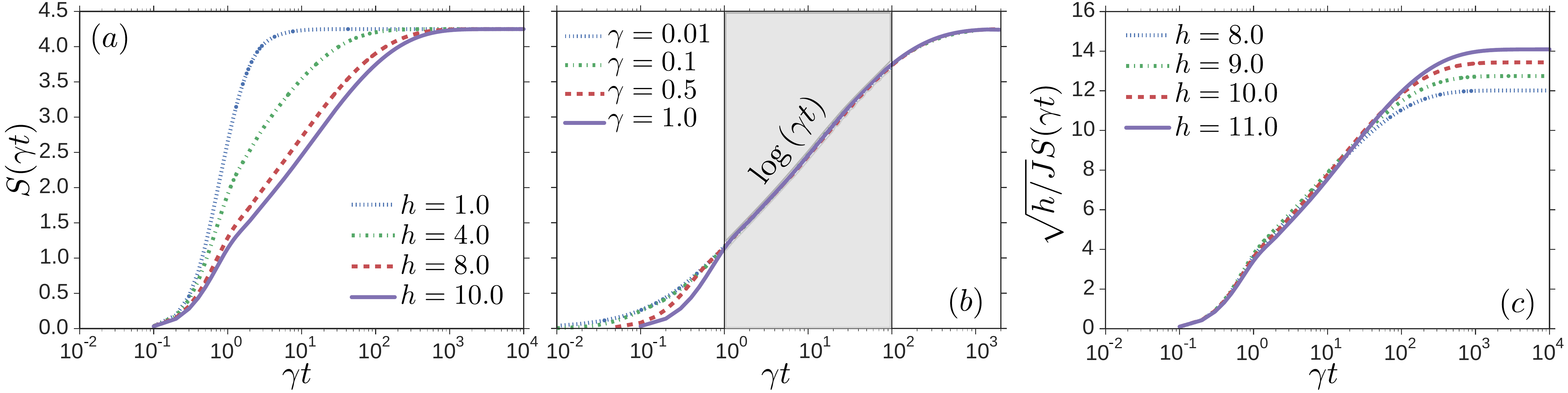}
\caption{Growth of von Neumann entropy $S$ in time from the initial separable state $\ket{\psi_0}$ (see text). Data are taken for a chain of $N=8$ sites, and $V=J=1.0$. The standard error (SE) is always smaller than the line width.
Panel (a) shows the different approaches of the entropy to equilibrium plotted as a function of $\gamma t$ for $\gamma=1.0$. In the ergodic phase the entropy displays a fast approach to its steady value, while in the MBL phase the approach is logarithmic. 
Panel (b) shows the entropy for different values of $\gamma$ at disorder $h=10.0$. The logarithmic window is shaded. 
In panel (c) we report the collapse of the entropy for various values of the disorder. Increasing $h$ leads to a stretching of the logarithmic phase.
}
\label{fig:1}
\end{figure*}

For Anderson insulators the coupling to low-temperature heat baths induces a nonzero, but highly suppressed, conductance in the context of variable-range hopping~\cite{Mott1969}. 
The Markovian bath considered in Eq.~(\ref{eq:defMasterEquation}), however, is fundamentally different in that it allows energy exchange at all scales, not just within a small low-temperature window, thereby providing a reservoir to overcome any desired energy mismatch. Still disorder is capable of increasing the lifetime of edge modes substantially~\cite{Carmele2015}. Moreover, Anderson-localized systems in the presence of dissipation display heterogenous relaxation dynamics~\cite{Genway2014}. The broadening of local spectra in MBL systems due to low-temperature baths has been studied recently~\cite{Nandkishore2014} and the persistence of MBL for small non-thermodynamic baths has been discussed in Refs.~\cite{Huse2015,Nandkishore2015a,Johri2015}. In \cite{Znidaric2010} it was shown that dephasing always induces diffusive transport in the steady-state of the considered system in its metallic phase.



\begin{figure*}[ht]
\centering
\includegraphics[trim = 0mm 0mm 0mm 0mm, clip, width=2.\columnwidth]{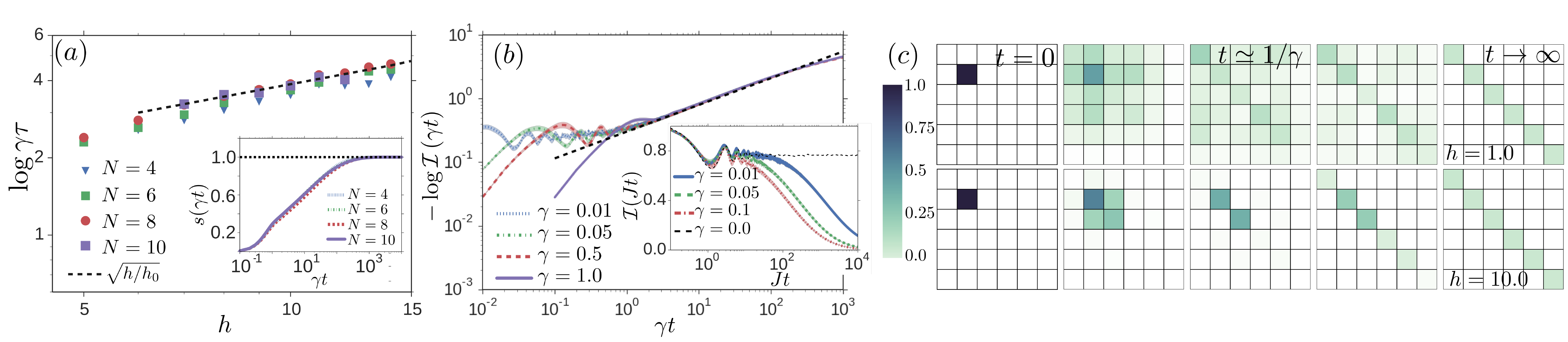}
\caption{(a) Scaling of the entropy for different system sizes, and $J=V=\gamma=1.0$ (data for $N=10$ are averaged over 20 random realizations). The SE is not shown as it is smaller than the data points. In the inset the normalized entropy density (see text) is plotted as a function of $\gamma t$. The main plot shows the divergence of the saturation time $\ts$ for increasing values of the disorder ($h$ axis in a logarithmic scale). The predicted behavior $\ts\sim \exp(\sqrt{h/h_0})$ is drawn for $h_0=1.5$ for comparison (black dashed line), and show significant agreement with the data in the MBL phase.
(b) Dynamics of the imbalance for $N=8$ and $J=V=1.0$. The SE is depicted as a shaded area around each curve. In the inset the imbalance is plotted for different values of decoherence rate as a function of $Jt$. For comparison the imbalance for the closed system is drawn as a dashed black line. In the main figure the asymptotic large time behavior $\mathcal{I}(\gamma t)\sim exp(-\mu t^\alpha)$ is tested, finding best fit parameters $\mu=0.3$ and $\alpha=0.42$ (see black dashed line).
(c) Qualitative sketch of the evolution of the density matrix. The density matrix is expressed in the Fock basis of the half-filling sector for $N=4$, and its elements $\left|\rho_{k,l}\right|$ are shown.
}
\label{fig:2}
\end{figure*}

\textit{Entropy.---} 
We quantify the heating induced by the bath via the von Neumann entropy $S(t) = -\left\langle \mathrm{Tr} \left[ \rs(t) \log(\rs(t)) \right]\right\rangle$ averaged over disorder realizations (where $k_B=1$). While the initial state is pure, dissipation will, by information transfer to the environment, lead to a mixed state of the fermionic chain with a nonzero entropy. In Fig.~\ref{fig:1}, we summarize our main results on $S(t)$ obtained by numerical integration of the Master equation (\ref{eq:defMasterEquation}) \cite{Qutip1,Qutip2} for systems up to $N=8$ sites and 100 realizations of the disorder.  (We emphasize already at this point that the finite-size dependence of the studied quantities is very weak, see also Fig.~\ref{fig:2}.)

For weak disorder, corresponding to the ergodic phase of the Hamiltonian $\Hs$ \cite{Pal2010,Luitz2015,Serbyn2015b}, the entropy saturates  quickly to its infinite-temperature value $S_\infty$.  In contrast, in the presence of strong disorder an extended temporal regime of slow growth emerges; see Fig.\ \ref{fig:1}(a).  
Remarkably, the entropy at weak system-bath coupling, $\gamma/J \lesssim 1$, exhibits a collapse for $t \gg \gamma^{-1}$ when rescaling the time axis by $\gamma$; see Fig.~\ref{fig:1}(b). Thus, for the investigated parameter regime, the entropy $S(t)$ depends on $\gamma$ only parametrically. In particular, the rate of the entropic growth, i.e. the heating itself, is independent of $\gamma$. Importantly, this rate is only set by the disorder strength. As we show in Fig.~\ref{fig:1}(c), rescaling $S(t)$ by $\sqrt{h/J}$, we find that the entropies for different $h$ collapse onto each other up to times where they start to saturate towards their steady state value $S_\infty$. Within our numerics we find that the respective growth is logarithmically slow as can be seen from Figs.~\ref{fig:1}(b) and (c). In particular, increasing the disorder strength, the temporal region of logarithmic dependence can be extended up to two decades, as for example in Fig.~\ref{fig:1}(c). 
As a consequence of this analysis, we conclude that $S(t)$ exhibits the following general form for times $t\gg \gamma^{-1}$ but before saturation
\begin{equation}
 \label{eq:universalfunc}
 S(t)\propto \sqrt{\frac{J}{h}} \log \left(\gamma t\right).
\end{equation}
This slow heating is remarkable: the Markovian bath can in principle provide any desired energy to overcome off-resonant hopping processes; however, our numerics suggest that these hopping processes are highly suppressed.

We now consider the dependence on $N$, see Fig.~\ref{fig:2}(a). The steady state  entropy $S_\infty$+ is easy to obtain: since the steady state is the identity, i.e. $\rs(t\to \infty)=\mathbb{1}$, $S_\infty=\log \mathcal{N}$ is determined by the total number of accessible states $\mathcal{N}$, with $\mathcal{N}=N!/[(N/2)!]^2$ at half-filling.  For large $N$, $S_\infty = N \log(2)$ as expected. This matches perfectly the numerically obtained values, see inset in Fig.~\ref{fig:2}(a) where we show the normalized entropy density, $s(t) = S(t)/S_\infty$, for different sizes $N$.  Overall, we find that the finite-size dependence is weak and affects only marginally the temporal evolution of the studied quantities including the entropy.
We attribute this weak dependence to the character of heating which occurs locally and not via the excitation of long-wavelength modes (see also the discussion at the end of the article).

The time for approaching stationarity becomes large in the strongly disordered regime due to the slow logarithmic growth.  We can extract a timescale $\ts$ without  assumptions on the functional form $S(t)$ from $\ts = \int_0^\infty \mathrm{d}t \,\, [1 - S(t) / S_\infty]$, which is finite as long as relaxation goes faster than $1/t$. The growth of $S(t)$ is logarithmic for intermediate times, but saturation to $S_\infty$ is much faster, leading to a finite $\ts$.  In Fig.~\ref{fig:2}(a) we show the obtained $\ts$ for different disorder strengths and system sizes. From the observed scaling in Fig.~\ref{fig:1}(c) it is possible to obtain an estimate of the relaxation time scale: 
Assuming the general form (\ref{eq:universalfunc}) for the entropy, we have that $\ts$ is set by the time where $S(\ts)\approx N \log(2)$, which gives
$\ts \sim \gamma^{-1} e^{\sqrt{h/h_0}}$,
with $h_0$ an energy scale which cannot be determined using this argument.  This behaviour is tested in Fig.~\ref{fig:2}(a).

\begin{figure*}[ht]
\centering
\includegraphics[trim = 0mm 0mm 0mm 0mm, clip, width=1.9\columnwidth]{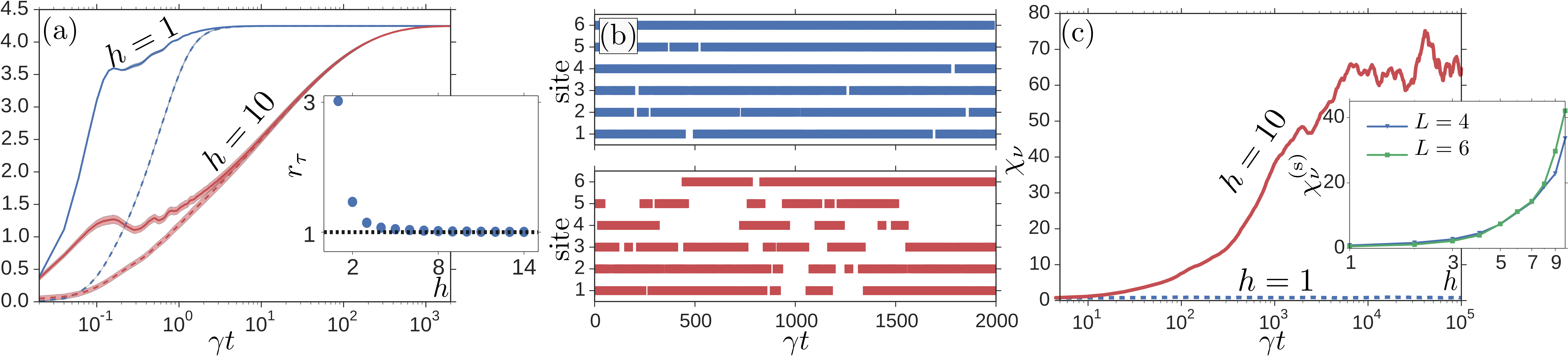}
\caption{(Colors online) All data are taken for $J=V=1$ and for $h=1$ (blue) and $h=10$ (red). In panel (a) we compare the diagonal entropy (solid lines) with the von Neumann entropy (dashed lines). All curves are plotted for $\gamma=0.2$, and the SE is depicted as a shaded area around each curve. In the inset the ratio $r_\tau$ is reported for different values of the disorder. Panel (b) shows an example of history of emissions for $\gamma=0.5$ in the evolution from the state $\ket{\psi_0}$ in a chain of six sites for one disorder realization. We consider $h=1.0$ in the top panel, and $h=10.0$ in the bottom panel. In panel (c) we show the dynamics of the emission susceptibility averaged over 100 repetitions of the experiment for one disorder realization. In the inset we report the dependence of the stationary value of the susceptibility on the disorder, averaged over up to 10 disorder realization and 100 trajectories.}

\label{fig:4}
\end{figure*}

\textit{Imbalance.---}
While the entropy quantifies heating in the system, we now study its ergodic properties.  In the MBL phase of $H$, in the absence of dissipation, failure to thermalise is evident in a strong memory of the initial conditions.  As has been done in recent experiments~\cite{Schreiber2015,Bordia2015,Smith2015} we quantify this memory via the imbalance
\be
	\mathcal{I}(t) = \frac{N_o(t)-N_e(t)}{N_o(t)+N_e(t)},
\ee
with $N_{o}$ ($N_{e}$) the number of fermions in odd (even) sites. Note that $\mathcal{I}(t)$ has an additional deep relation to a generalization of the Anderson localization length to many-body Hilbert space~\cite{Hauke2014}: $\mathcal{D}(t) = (N/2)[1-\mathcal{I}(t)]$ with $\mathcal{D}(t)$  quantifying the mean Hamming distance the system departs from its initial Fock state $|\psi_0\rangle$.

In Fig.~\ref{fig:2}(b) we show the imbalance for large disorder, such that the closed system is in the MBL phase.  We see from the inset that for times up to $\gamma^{-1}$, the imbalance closely follows the coherent evolution.  In contrast to the closed nonergodic system where $\mathcal{I}(t \to \infty) > 0$, in the presence dissipation the system loses memory of the initial state for times $t>\gamma^{-1}$. We find that the long-time behavior is approximately a stretched exponential
\be
	\mathcal{I}(t) \stackrel{t\to\infty}{\longrightarrow} e^{-\mu t^{\alpha}},
\ee
thus slower than exponential (which is characteristic of time-correlators in systems with slow relaxation \cite{Binder2011}). Remarkably, the dependence of the imbalance on the system-bath coupling $\gamma$ is again only parametric as it can be absorbed  into a rescaling of the time. At the longest times simulated, $t \gtrsim 10^2 \gamma^{-1}$, one sees deviation from the stretched exponential scaling. Whether this is a consequence of the finite size of our system we cannot address on the basis of our methodology. 

\textit{Connection to experiment and in situ monitoring of MBL.---} The physics discussed here is observable in experiments on lattice gases of ultracold atoms, see e.g.\ Ref.\ \cite{Schreiber2015} where MBL was explored with fermions in an instance of the Aubry-Andr\'{e} model. Due to the robustness of MBL phenomena we expect that with dissipation the physics will be similar to our simplified model, Eqs.\ (\ref{eq:defHamiltonian}-\ref{eq:defMasterEquation}), which can be thought of as a Fermi-Hubbard model with long-range interactions (see e.g. Ref. \cite{Dutta2015}). 

In current experiments~\cite{Schreiber2015,Bordia2015,Smith2015} a readout of the positions of the excitations is usually performed, giving access to the diagonal elements of the density matrix (in the Fock basis). The diagonal entropy $S_d(t)=\sum_n \rho_{nn}(t) \log \rho_{nn}(t)$ is then computable and gives an upper bound for $S(t)$~\cite{Polkovnikov2011}. Notice that the generic basis chosen for $S_d(t)$ is the basis of eigenstates of the Hamiltonian. Importantly, in the regime of strong disorder this eigenbasis is close to the Fock basis~\cite{Serbyn2013}. This gives a  good estimate of the true entropy $S(t)$, see Fig.~\ref{fig:4}(a). In particular, 
in the inset we compare the time scale $\tau_d$ obtained with $S_d(t)$ with $\tau$ obtained above, cf.\ Fig.\ \ref{fig:2}(a), through the ratio $r_\tau=\tau/\tau_d$. We find $r_\tau \approx 1$ in the MBL regime. This can be understood from the sketch of the evolution of $\rho(t)$ in the Fock basis shown in Fig.~\ref{fig:2}(c). Coherences are established due to $H$ over times $t\leq 1/\gamma$: For small $h$ such a superposition is over many Fock states, subsequently transformed into a mixture rapidly approaching $\rs(t\to \infty)=\mathbb{1}$. For large $h$, in contrast, $H$  is inefficient in creating large superpositions, and the resultant diagonal density matrix due to dephasing is far from the identity. Subsequent evolution can be regarded as happening only on the diagonal elements, gradually mixing towards the eventual $\rs(t\to \infty)=\mathbb{1}$. This difference in the spreading of the superposition formed by $H$ is responsible for how well $S(t)$ can be approximated with $S_d(t)$.

Importantly, dissipation as introduced in Eq.\ (\ref{eq:defMasterEquation}) emerges naturally in cold atom lattice experiments. The underlying mechanism is that of off-resonant scattering of photons from the laser field that forms the lattice trapping potential. As shown in Ref. \cite{Sarkar2014} this leads directly to the dissipator in Eq.(\ref{eq:defMasterEquation}) when considering the dynamics of fermions in the lowest band of the lattice. The rate $\gamma$ is then controlled by the detuning and strength of the trapping laser.
In principle, it is possible to record scattered photons giving the opportunity for observing MBL related phenomena in situ. 
This can be efficiently simulated via quantum jump Monte Carlo \cite{Qjumps1,Qjumps2,Daley2014}. 
In Fig.\ \ref{fig:4}(b) we show the site-resolved photon signal from simulated quantum jump trajectories. The MBL phase is characterized by intermittent emissions. Although the dissipation acts with the same rate on each site the probability of an emission is proportional to the occupation probability.
On the MBL side, occupation probability is strongly inhomogeneous due to the disorder, and its dynamics is slow. The dissipator in Eq.\ (\ref{eq:defMasterEquation}) is projective on the occupation basis causing long periods of repeated emissions, alternating with long periods of no emissions.
This ``dynamic heterogeneity'' in quantum jump trajectories is reminiscent of classical glassy systems \cite{Garrahan2002}, and can be quantified by means of a dynamical susceptibility $\chi_\nu(t)$: if $\nu_i(t)$ denotes the accumulated number of emissions on site $i$ up to time $t$, we define $\chi_\nu(t)=\left\langle\sum_{i} \left[ \nu_i(t)-\langle\nu_i(t)\rangle \right]^2 \right\rangle / (Nt)$, where the mean is taken over many trajectories and for a fixed realisation of the disorder.  
While $\chi_\nu(t)$ quickly saturates in the ergodic phase, on the MBL side it grows slowly as indicative of pronounced time correlations in the emissions, see Fig.\ \ref{fig:4}(c).   
The saturation value $\chi_\nu^{\mathrm{(s)}}$ displays a very different behavior on the ergodic side where it is close to zero, and the MBL side, where it shows a rapid increase with the disorder strength (see inset).
It is important to note that the overall emission rate is the same in the ergodic and MBL sides: the rate of emissions is controlled by $H_{\rm eff} = H - i\frac{\gamma}{2} \sum_{i} n_{i}$, and since density is conserved overall emissions are simply a Poisson process with constant rate $\gamma N/2$, and therefore independent of $H$.  It is the spatial pattern of emissions, as in Fig.\ \ref{fig:4}(b), that is sensitive to $H$ and gives a clear indication of the slow evolution and long correlation times associated to MBL dynamics.

\textit{Discussion.---} 
The physical picture for the observed slow heating is the following.  For strong disorder, $H$ of Eq.~(\ref{eq:defHamiltonian}) is known to have a representation
$H = \sum_{l} \varepsilon_l \nu_l + \sum_{lm} K_{lm} \nu_l \nu_m + \dots$
in terms of local integrals of motion $\nu_l=Z_l n_l+\mathcal{O}(J/W,U/W)$ perturbatively connected to the bare occupacies $n_l$, where $Z_l \lesssim 1$, and energies $\varepsilon_l$ and couplings $K_{lm}$ are functions of the parameters of the original Hamiltonian \cite{Nandkishore2015}. In this regime, the jump operators $n_l$ in the Master equation (\ref{eq:defMasterEquation}) almost commute with the local integrals of motion, $[n_l,\nu_l] = \mathcal{O}(J/W,U/W)$, the influence of the bath is weak, and at infinite disorder heating is completely absent.  This is indeed seen in the emission behaviour of individual trajectories, cf.\ Fig.\ \ref{fig:4}.   Furthermore, in this regime and in the weak coupling limit, $\gamma \to 0$, quantum jumps can be considered independent such that between two jumps the free evolution dominantly leads to dephasing, making the density matrix approximately diagonal in the basis of the local integrals of motion. A quantum jump then only slightly perturbs the state of the system because projecting onto local occupation numbers does not fully project onto the local integrals of motion. This slight mismatch between local integrals of motion and local occupations implies that the quantum jumps only weakly perturb the system and therefore only slightly increase the system's energy. This then leads to the slow heating observed numerically.

\textit{Acknowledgements.} --- E.L.\ would like to thank M.\ Marcuzzi and M.\ van Horssen for useful discussions. M.H. acknowledges valuable discussions with E. Altman, I. Bloch, P. Bordia, H. L\"uschen, and M. Schreiber. This work was supported by European Research Council under the European Unions Seventh Framework Programme (FP/2007-2013) / ERC Grant Agreement No. 335266 (ESCQUMA), by EPSRC Grant No.\ EP/J009776/1, and the Deutsche Akademie der Naturforscher Leopoldina (grant No. LPDS 2013-07 and LPDR 2015-01).

\bibliographystyle{apsrev4-1}


\begin{thebibliography}{53}%
\makeatletter
\providecommand \@ifxundefined [1]{%
 \@ifx{#1\undefined}
}%
\providecommand \@ifnum [1]{%
 \ifnum #1\expandafter \@firstoftwo
 \else \expandafter \@secondoftwo
 \fi
}%
\providecommand \@ifx [1]{%
 \ifx #1\expandafter \@firstoftwo
 \else \expandafter \@secondoftwo
 \fi
}%
\providecommand \natexlab [1]{#1}%
\providecommand \enquote  [1]{``#1''}%
\providecommand \bibnamefont  [1]{#1}%
\providecommand \bibfnamefont [1]{#1}%
\providecommand \citenamefont [1]{#1}%
\providecommand \href@noop [0]{\@secondoftwo}%
\providecommand \href [0]{\begingroup \@sanitize@url \@href}%
\providecommand \@href[1]{\@@startlink{#1}\@@href}%
\providecommand \@@href[1]{\endgroup#1\@@endlink}%
\providecommand \@sanitize@url [0]{\catcode `\\12\catcode `\$12\catcode
  `\&12\catcode `\#12\catcode `\^12\catcode `\_12\catcode `\%12\relax}%
\providecommand \@@startlink[1]{}%
\providecommand \@@endlink[0]{}%
\providecommand \url  [0]{\begingroup\@sanitize@url \@url }%
\providecommand \@url [1]{\endgroup\@href {#1}{\urlprefix }}%
\providecommand \urlprefix  [0]{URL }%
\providecommand \Eprint [0]{\href }%
\providecommand \doibase [0]{http://dx.doi.org/}%
\providecommand \selectlanguage [0]{\@gobble}%
\providecommand \bibinfo  [0]{\@secondoftwo}%
\providecommand \bibfield  [0]{\@secondoftwo}%
\providecommand \translation [1]{[#1]}%
\providecommand \BibitemOpen [0]{}%
\providecommand \bibitemStop [0]{}%
\providecommand \bibitemNoStop [0]{.\EOS\space}%
\providecommand \EOS [0]{\spacefactor3000\relax}%
\providecommand \BibitemShut  [1]{\csname bibitem#1\endcsname}%
\let\auto@bib@innerbib\@empty
\bibitem [{\citenamefont {Peliti}(2011)}]{Peliti2011}%
  \BibitemOpen
  \bibfield  {author} {\bibinfo {author} {\bibfnamefont {L.}~\bibnamefont
  {Peliti}},\ }\href@noop {} {\emph {\bibinfo {title} {{Statistical mechanics
  in a nutshell}}}}\ (\bibinfo  {publisher} {Princeton University Press},\
  \bibinfo {year} {2011})\BibitemShut {NoStop}%
\bibitem [{\citenamefont {Eisert}\ \emph {et~al.}(2014)\citenamefont {Eisert},
  \citenamefont {Friesdorf},\ and\ \citenamefont {Gogolin}}]{Eisert2014}%
  \BibitemOpen
  \bibfield  {author} {\bibinfo {author} {\bibfnamefont {J.}~\bibnamefont
  {Eisert}}, \bibinfo {author} {\bibfnamefont {M.}~\bibnamefont {Friesdorf}}, \
  and\ \bibinfo {author} {\bibfnamefont {C.}~\bibnamefont {Gogolin}},\ }\href
  {\doibase 10.1038/NPHYS3215} {\bibfield  {journal} {\bibinfo  {journal}
  {Nature Physics}\ }\textbf {\bibinfo {volume} {11}},\ \bibinfo {pages} {7}
  (\bibinfo {year} {2014})}\BibitemShut {NoStop}%
\bibitem [{\citenamefont {D'Alessio}\ \emph {et~al.}()\citenamefont
  {D'Alessio}, \citenamefont {Kafri}, \citenamefont {Polkovnikov},\ and\
  \citenamefont {Rigol}}]{Dalessio2015}%
  \BibitemOpen
  \bibfield  {author} {\bibinfo {author} {\bibfnamefont {L.}~\bibnamefont
  {D'Alessio}}, \bibinfo {author} {\bibfnamefont {Y.}~\bibnamefont {Kafri}},
  \bibinfo {author} {\bibfnamefont {A.}~\bibnamefont {Polkovnikov}}, \ and\
  \bibinfo {author} {\bibfnamefont {M.}~\bibnamefont {Rigol}},\ }\href@noop {}
  {\bibinfo  {journal} {arXiv:1509.06411}\ }\BibitemShut {NoStop}%
\bibitem [{\citenamefont {Binder}\ and\ \citenamefont
  {Kob}(2011)}]{Binder2011}%
  \BibitemOpen
\bibfield  {journal} {  }\bibfield  {author} {\bibinfo {author} {\bibfnamefont
  {K.}~\bibnamefont {Binder}}\ and\ \bibinfo {author} {\bibfnamefont
  {W.}~\bibnamefont {Kob}},\ }\href@noop {} {\emph {\bibinfo {title} {{Glassy
  Materials and Disordered Solids}}}}\ (\bibinfo  {publisher} {World
  Scientific},\ \bibinfo {year} {2011})\BibitemShut {NoStop}%
\bibitem [{\citenamefont {Biroli}\ and\ \citenamefont
  {Garrahan}(2013)}]{Biroli2013}%
  \BibitemOpen
  \bibfield  {author} {\bibinfo {author} {\bibfnamefont {G.}~\bibnamefont
  {Biroli}}\ and\ \bibinfo {author} {\bibfnamefont {J.~P.}\ \bibnamefont
  {Garrahan}},\ }\href {\doibase 10.1063/1.4795539} {\bibfield  {journal}
  {\bibinfo  {journal} {J. Chem. Phys.}\ }\textbf {\bibinfo {volume} {138}},\
  \bibinfo {eid} {12A301} (\bibinfo {year} {2013})}\BibitemShut {NoStop}%
\bibitem [{\citenamefont {Altshuler}\ \emph {et~al.}(1997)\citenamefont
  {Altshuler}, \citenamefont {Gefen}, \citenamefont {Kamenev},\ and\
  \citenamefont {Levitov}}]{Altshuler1997}%
  \BibitemOpen
  \bibfield  {author} {\bibinfo {author} {\bibfnamefont {B.~L.}\ \bibnamefont
  {Altshuler}}, \bibinfo {author} {\bibfnamefont {Y.}~\bibnamefont {Gefen}},
  \bibinfo {author} {\bibfnamefont {A.}~\bibnamefont {Kamenev}}, \ and\
  \bibinfo {author} {\bibfnamefont {L.~S.}\ \bibnamefont {Levitov}},\
  }\href@noop {} {\bibfield  {journal} {\bibinfo  {journal} {Phys. Rev. Lett.}\
  }\textbf {\bibinfo {volume} {78}},\ \bibinfo {pages} {2803} (\bibinfo {year}
  {1997})}\BibitemShut {NoStop}%
\bibitem [{\citenamefont {Basko}\ \emph {et~al.}(2006)\citenamefont {Basko},
  \citenamefont {Aleiner},\ and\ \citenamefont {Altshuler}}]{Basko2006}%
  \BibitemOpen
  \bibfield  {author} {\bibinfo {author} {\bibfnamefont {D.}~\bibnamefont
  {Basko}}, \bibinfo {author} {\bibfnamefont {I.}~\bibnamefont {Aleiner}}, \
  and\ \bibinfo {author} {\bibfnamefont {B.}~\bibnamefont {Altshuler}},\
  }\href@noop {} {\bibfield  {journal} {\bibinfo  {journal} {Ann. of Phys.}\
  }\textbf {\bibinfo {volume} {321}},\ \bibinfo {pages} {1126} (\bibinfo {year}
  {2006})}\BibitemShut {NoStop}%
\bibitem [{\citenamefont {Gornyi}\ \emph {et~al.}(2005)\citenamefont {Gornyi},
  \citenamefont {Mirlin},\ and\ \citenamefont {Polyakov}}]{Gornyi2005}%
  \BibitemOpen
  \bibfield  {author} {\bibinfo {author} {\bibfnamefont {I.}~\bibnamefont
  {Gornyi}}, \bibinfo {author} {\bibfnamefont {A.}~\bibnamefont {Mirlin}}, \
  and\ \bibinfo {author} {\bibfnamefont {D.}~\bibnamefont {Polyakov}},\ }\href
  {\doibase 10.1103/PhysRevLett.95.206603} {\bibfield  {journal} {\bibinfo
  {journal} {Phys. Rev. Lett.}\ }\textbf {\bibinfo {volume} {95}},\ \bibinfo
  {pages} {206603} (\bibinfo {year} {2005})}\BibitemShut {NoStop}%
\bibitem [{\citenamefont {Oganesyan}\ and\ \citenamefont
  {Huse}(2007)}]{Oganesyan2007}%
  \BibitemOpen
  \bibfield  {author} {\bibinfo {author} {\bibfnamefont {V.}~\bibnamefont
  {Oganesyan}}\ and\ \bibinfo {author} {\bibfnamefont {D.~A.}\ \bibnamefont
  {Huse}},\ }\href {\doibase 10.1103/PhysRevB.75.155111} {\bibfield  {journal}
  {\bibinfo  {journal} {Phys. Rev. B}\ }\textbf {\bibinfo {volume} {75}},\
  \bibinfo {pages} {155111} (\bibinfo {year} {2007})}\BibitemShut {NoStop}%
\bibitem [{\citenamefont {Pal}\ and\ \citenamefont {Huse}(2010)}]{Pal2010}%
  \BibitemOpen
  \bibfield  {author} {\bibinfo {author} {\bibfnamefont {A.}~\bibnamefont
  {Pal}}\ and\ \bibinfo {author} {\bibfnamefont {D.~A.}\ \bibnamefont {Huse}},\
  }\href@noop {} {\bibfield  {journal} {\bibinfo  {journal} {Phys. Rev. B}\
  }\textbf {\bibinfo {volume} {82}},\ \bibinfo {pages} {174411} (\bibinfo
  {year} {2010})}\BibitemShut {NoStop}%
\bibitem [{\citenamefont {Canovi}\ \emph {et~al.}(2011)\citenamefont {Canovi},
  \citenamefont {Rossini}, \citenamefont {Fazio}, \citenamefont {Santoro},\
  and\ \citenamefont {Silva}}]{Canovi2011}%
  \BibitemOpen
  \bibfield  {author} {\bibinfo {author} {\bibfnamefont {E.}~\bibnamefont
  {Canovi}}, \bibinfo {author} {\bibfnamefont {D.}~\bibnamefont {Rossini}},
  \bibinfo {author} {\bibfnamefont {R.}~\bibnamefont {Fazio}}, \bibinfo
  {author} {\bibfnamefont {G.~E.}\ \bibnamefont {Santoro}}, \ and\ \bibinfo
  {author} {\bibfnamefont {A.}~\bibnamefont {Silva}},\ }\href {\doibase
  10.1103/PhysRevB.83.094431} {\bibfield  {journal} {\bibinfo  {journal} {Phys.
  Rev. B}\ }\textbf {\bibinfo {volume} {83}},\ \bibinfo {pages} {094431}
  (\bibinfo {year} {2011})}\BibitemShut {NoStop}%
\bibitem [{\citenamefont {Bardarson}\ \emph {et~al.}(2012)\citenamefont
  {Bardarson}, \citenamefont {Pollmann},\ and\ \citenamefont
  {Moore}}]{Bardarson2012}%
  \BibitemOpen
  \bibfield  {author} {\bibinfo {author} {\bibfnamefont {J.~H.}\ \bibnamefont
  {Bardarson}}, \bibinfo {author} {\bibfnamefont {F.}~\bibnamefont {Pollmann}},
  \ and\ \bibinfo {author} {\bibfnamefont {J.~E.}\ \bibnamefont {Moore}},\
  }\href {\doibase 10.1103/PhysRevLett.109.017202} {\bibfield  {journal}
  {\bibinfo  {journal} {Phys. Rev. Lett.}\ }\textbf {\bibinfo {volume} {109}},\
  \bibinfo {pages} {017202} (\bibinfo {year} {2012})}\BibitemShut {NoStop}%
\bibitem [{\citenamefont {Serbyn}\ \emph {et~al.}(2013)\citenamefont {Serbyn},
  \citenamefont {{Papi\ifmmode \acute{c}\else {\'c}\fi{}}},\ and\ \citenamefont
  {Abanin}}]{Serbyn2013}%
  \BibitemOpen
  \bibfield  {author} {\bibinfo {author} {\bibfnamefont {M.}~\bibnamefont
  {Serbyn}}, \bibinfo {author} {\bibfnamefont {Z.}~\bibnamefont {{Papi\ifmmode
  \acute{c}\else {\'c}\fi{}}}}, \ and\ \bibinfo {author} {\bibfnamefont
  {D.~A.}\ \bibnamefont {Abanin}},\ }\href {\doibase
  10.1103/PhysRevLett.110.260601} {\bibfield  {journal} {\bibinfo  {journal}
  {Phys. Rev. Lett.}\ }\textbf {\bibinfo {volume} {110}},\ \bibinfo {pages}
  {260601} (\bibinfo {year} {2013})}\BibitemShut {NoStop}%
\bibitem [{\citenamefont {Huse}\ \emph {et~al.}(2014)\citenamefont {Huse},
  \citenamefont {Nandkishore},\ and\ \citenamefont {Oganesyan}}]{Huse2014}%
  \BibitemOpen
  \bibfield  {author} {\bibinfo {author} {\bibfnamefont {D.~A.}\ \bibnamefont
  {Huse}}, \bibinfo {author} {\bibfnamefont {R.}~\bibnamefont {Nandkishore}}, \
  and\ \bibinfo {author} {\bibfnamefont {V.}~\bibnamefont {Oganesyan}},\ }\href
  {\doibase 10.1103/PhysRevB.90.174202} {\bibfield  {journal} {\bibinfo
  {journal} {Phys. Rev. B}\ }\textbf {\bibinfo {volume} {90}},\ \bibinfo
  {pages} {174202} (\bibinfo {year} {2014})}\BibitemShut {NoStop}%
\bibitem [{\citenamefont {Imbrie}()}]{Imbrie2014}%
  \BibitemOpen
  \bibfield  {author} {\bibinfo {author} {\bibfnamefont {J.~Z.}\ \bibnamefont
  {Imbrie}},\ }\href@noop {} {\ }\Eprint {http://arxiv.org/abs/1403.7837}
  {arXiv:1403.7837} \BibitemShut {NoStop}%
\bibitem [{\citenamefont {Ros}\ \emph {et~al.}(2015)\citenamefont {Ros},
  \citenamefont {M{\"u}ller},\ and\ \citenamefont
  {Scardicchio}}]{Scardicchio2015}%
  \BibitemOpen
  \bibfield  {author} {\bibinfo {author} {\bibfnamefont {V.}~\bibnamefont
  {Ros}}, \bibinfo {author} {\bibfnamefont {M.}~\bibnamefont {M{\"u}ller}}, \
  and\ \bibinfo {author} {\bibfnamefont {A.}~\bibnamefont {Scardicchio}},\
  }\href {\doibase http://dx.doi.org/10.1016/j.nuclphysb.2014.12.014}
  {\bibfield  {journal} {\bibinfo  {journal} {Nuclear Physics B}\ }\textbf
  {\bibinfo {volume} {891}},\ \bibinfo {pages} {420 } (\bibinfo {year}
  {2015})}\BibitemShut {NoStop}%
\bibitem [{\citenamefont {Andraschko}\ \emph {et~al.}(2014)\citenamefont
  {Andraschko}, \citenamefont {Enss},\ and\ \citenamefont
  {Sirker}}]{Andraschko2014}%
  \BibitemOpen
  \bibfield  {author} {\bibinfo {author} {\bibfnamefont {F.}~\bibnamefont
  {Andraschko}}, \bibinfo {author} {\bibfnamefont {T.}~\bibnamefont {Enss}}, \
  and\ \bibinfo {author} {\bibfnamefont {J.}~\bibnamefont {Sirker}},\ }\href
  {\doibase 10.1103/PhysRevLett.113.217201} {\bibfield  {journal} {\bibinfo
  {journal} {Phys. Rev. Lett.}\ }\textbf {\bibinfo {volume} {113}},\ \bibinfo
  {pages} {217201} (\bibinfo {year} {2014})}\BibitemShut {NoStop}%
\bibitem [{\citenamefont {Yao}\ \emph {et~al.}(2014)\citenamefont {Yao},
  \citenamefont {Laumann}, \citenamefont {Gopalakrishnan}, \citenamefont
  {Knap}, \citenamefont {M{\"u}ller}, \citenamefont {Demler},\ and\
  \citenamefont {Lukin}}]{Yao2014}%
  \BibitemOpen
  \bibfield  {author} {\bibinfo {author} {\bibfnamefont {N.~Y.}\ \bibnamefont
  {Yao}}, \bibinfo {author} {\bibfnamefont {C.~R.}\ \bibnamefont {Laumann}},
  \bibinfo {author} {\bibfnamefont {S.}~\bibnamefont {Gopalakrishnan}},
  \bibinfo {author} {\bibfnamefont {M.}~\bibnamefont {Knap}}, \bibinfo {author}
  {\bibfnamefont {M.}~\bibnamefont {M{\"u}ller}}, \bibinfo {author}
  {\bibfnamefont {E.~A.}\ \bibnamefont {Demler}}, \ and\ \bibinfo {author}
  {\bibfnamefont {M.~D.}\ \bibnamefont {Lukin}},\ }\href {\doibase
  10.1103/PhysRevLett.113.243002} {\bibfield  {journal} {\bibinfo  {journal}
  {Phys. Rev. Lett.}\ }\textbf {\bibinfo {volume} {113}},\ \bibinfo {pages}
  {243002} (\bibinfo {year} {2014})}\BibitemShut {NoStop}%
\bibitem [{\citenamefont {Serbyn}\ \emph {et~al.}(2014)\citenamefont {Serbyn},
  \citenamefont {{Papi\ifmmode \acute{c}\else {\'c}\fi{}}},\ and\ \citenamefont
  {Abanin}}]{Serbyn2014}%
  \BibitemOpen
  \bibfield  {author} {\bibinfo {author} {\bibfnamefont {M.}~\bibnamefont
  {Serbyn}}, \bibinfo {author} {\bibfnamefont {Z.}~\bibnamefont {{Papi\ifmmode
  \acute{c}\else {\'c}\fi{}}}}, \ and\ \bibinfo {author} {\bibfnamefont
  {D.~A.}\ \bibnamefont {Abanin}},\ }\href {\doibase
  10.1103/PhysRevB.90.174302} {\bibfield  {journal} {\bibinfo  {journal} {Phys.
  Rev. B}\ }\textbf {\bibinfo {volume} {90}},\ \bibinfo {pages} {174302}
  (\bibinfo {year} {2014})}\BibitemShut {NoStop}%
\bibitem [{\citenamefont {Vasseur}\ \emph {et~al.}(2015)\citenamefont
  {Vasseur}, \citenamefont {Parameswaran},\ and\ \citenamefont
  {Moore}}]{Vasseur2015}%
  \BibitemOpen
  \bibfield  {author} {\bibinfo {author} {\bibfnamefont {R.}~\bibnamefont
  {Vasseur}}, \bibinfo {author} {\bibfnamefont {S.~A.}\ \bibnamefont
  {Parameswaran}}, \ and\ \bibinfo {author} {\bibfnamefont {J.~E.}\
  \bibnamefont {Moore}},\ }\href {\doibase 10.1103/PhysRevB.91.140202}
  {\bibfield  {journal} {\bibinfo  {journal} {Phys. Rev. B}\ }\textbf {\bibinfo
  {volume} {91}},\ \bibinfo {pages} {140202} (\bibinfo {year}
  {2015})}\BibitemShut {NoStop}%
\bibitem [{\citenamefont {Agarwal}\ \emph {et~al.}(2015)\citenamefont
  {Agarwal}, \citenamefont {Gopalakrishnan}, \citenamefont {Knap},
  \citenamefont {M{\"u}ller},\ and\ \citenamefont {Demler}}]{Agarwal2015}%
  \BibitemOpen
  \bibfield  {author} {\bibinfo {author} {\bibfnamefont {K.}~\bibnamefont
  {Agarwal}}, \bibinfo {author} {\bibfnamefont {S.}~\bibnamefont
  {Gopalakrishnan}}, \bibinfo {author} {\bibfnamefont {M.}~\bibnamefont
  {Knap}}, \bibinfo {author} {\bibfnamefont {M.}~\bibnamefont {M{\"u}ller}}, \
  and\ \bibinfo {author} {\bibfnamefont {E.}~\bibnamefont {Demler}},\ }\href
  {\doibase 10.1103/PhysRevLett.114.160401} {\bibfield  {journal} {\bibinfo
  {journal} {Phys. Rev. Lett.}\ }\textbf {\bibinfo {volume} {114}},\ \bibinfo
  {pages} {160401} (\bibinfo {year} {2015})}\BibitemShut {NoStop}%
\bibitem [{\citenamefont {Laumann}\ \emph {et~al.}(2014)\citenamefont
  {Laumann}, \citenamefont {Pal},\ and\ \citenamefont
  {Scardicchio}}]{Laumann2014}%
  \BibitemOpen
  \bibfield  {author} {\bibinfo {author} {\bibfnamefont {C.~R.}\ \bibnamefont
  {Laumann}}, \bibinfo {author} {\bibfnamefont {A.}~\bibnamefont {Pal}}, \ and\
  \bibinfo {author} {\bibfnamefont {A.}~\bibnamefont {Scardicchio}},\ }\href
  {\doibase 10.1103/PhysRevLett.113.200405} {\bibfield  {journal} {\bibinfo
  {journal} {Phys. Rev. Lett.}\ }\textbf {\bibinfo {volume} {113}},\ \bibinfo
  {pages} {200405} (\bibinfo {year} {2014})}\BibitemShut {NoStop}%
\bibitem [{\citenamefont {{De Roeck}}\ and\ \citenamefont
  {Huveneers}(2014)}]{De-Roeck2014}%
  \BibitemOpen
  \bibfield  {author} {\bibinfo {author} {\bibfnamefont {W.}~\bibnamefont {{De
  Roeck}}}\ and\ \bibinfo {author} {\bibfnamefont {F.~m.~c.}\ \bibnamefont
  {Huveneers}},\ }\href {\doibase 10.1103/PhysRevB.90.165137} {\bibfield
  {journal} {\bibinfo  {journal} {Phys. Rev. B}\ }\textbf {\bibinfo {volume}
  {90}},\ \bibinfo {pages} {165137} (\bibinfo {year} {2014})}\BibitemShut
  {NoStop}%
\bibitem [{\citenamefont {{Yao}}\ \emph {et~al.}()\citenamefont {{Yao}},
  \citenamefont {{Laumann}}, \citenamefont {{Cirac}}, \citenamefont {{Lukin}},\
  and\ \citenamefont {{Moore}}}]{Yao2014b}%
  \BibitemOpen
  \bibfield  {author} {\bibinfo {author} {\bibfnamefont {N.~Y.}\ \bibnamefont
  {{Yao}}}, \bibinfo {author} {\bibfnamefont {C.~R.}\ \bibnamefont
  {{Laumann}}}, \bibinfo {author} {\bibfnamefont {J.~I.}\ \bibnamefont
  {{Cirac}}}, \bibinfo {author} {\bibfnamefont {M.~D.}\ \bibnamefont
  {{Lukin}}}, \ and\ \bibinfo {author} {\bibfnamefont {J.~E.}\ \bibnamefont
  {{Moore}}},\ }\href@noop {} {\ }\Eprint {http://arxiv.org/abs/1410.7407}
  {arXiv:1410.7407} \BibitemShut {NoStop}%
\bibitem [{\citenamefont {van Horssen}\ \emph {et~al.}(2015)\citenamefont {van
  Horssen}, \citenamefont {Levi},\ and\ \citenamefont
  {Garrahan}}]{Horssen2015}%
  \BibitemOpen
  \bibfield  {author} {\bibinfo {author} {\bibfnamefont {M.}~\bibnamefont {van
  Horssen}}, \bibinfo {author} {\bibfnamefont {E.}~\bibnamefont {Levi}}, \ and\
  \bibinfo {author} {\bibfnamefont {J.~P.}\ \bibnamefont {Garrahan}},\ }\href
  {\doibase 10.1103/PhysRevB.92.100305} {\bibfield  {journal} {\bibinfo
  {journal} {Phys. Rev. B}\ }\textbf {\bibinfo {volume} {92}},\ \bibinfo
  {pages} {100305} (\bibinfo {year} {2015})}\BibitemShut {NoStop}%
\bibitem [{\citenamefont {Altman}\ and\ \citenamefont
  {Vosk}(2015)}]{Altman2015}%
  \BibitemOpen
  \bibfield  {author} {\bibinfo {author} {\bibfnamefont {E.}~\bibnamefont
  {Altman}}\ and\ \bibinfo {author} {\bibfnamefont {R.}~\bibnamefont {Vosk}},\
  }\href@noop {} {\bibfield  {journal} {\bibinfo  {journal} {Annu. Rev.
  Condens. Matter Phys.}\ }\textbf {\bibinfo {volume} {6}},\ \bibinfo {pages}
  {383} (\bibinfo {year} {2015})}\BibitemShut {NoStop}%
\bibitem [{\citenamefont {Nandkishore}\ and\ \citenamefont
  {Huse}(2015)}]{Nandkishore2015}%
  \BibitemOpen
  \bibfield  {author} {\bibinfo {author} {\bibfnamefont {R.}~\bibnamefont
  {Nandkishore}}\ and\ \bibinfo {author} {\bibfnamefont {D.~A.}\ \bibnamefont
  {Huse}},\ }\href@noop {} {\bibfield  {journal} {\bibinfo  {journal} {Annu.
  Rev. Condens. Matter Phys.}\ }\textbf {\bibinfo {volume} {6}},\ \bibinfo
  {pages} {15} (\bibinfo {year} {2015})}\BibitemShut {NoStop}%
\bibitem [{\citenamefont {Schreiber}\ \emph {et~al.}(2015)\citenamefont
  {Schreiber}, \citenamefont {Hodgman}, \citenamefont {Bordia}, \citenamefont
  {L{\~A}¼schen}, \citenamefont {Fischer}, \citenamefont {Vosk}, \citenamefont
  {Altman}, \citenamefont {Schneider},\ and\ \citenamefont
  {Bloch}}]{Schreiber2015}%
  \BibitemOpen
  \bibfield  {author} {\bibinfo {author} {\bibfnamefont {M.}~\bibnamefont
  {Schreiber}}, \bibinfo {author} {\bibfnamefont {S.~S.}\ \bibnamefont
  {Hodgman}}, \bibinfo {author} {\bibfnamefont {P.}~\bibnamefont {Bordia}},
  \bibinfo {author} {\bibfnamefont {H.~P.}\ \bibnamefont {L{\~A}¼schen}},
  \bibinfo {author} {\bibfnamefont {M.~H.}\ \bibnamefont {Fischer}}, \bibinfo
  {author} {\bibfnamefont {R.}~\bibnamefont {Vosk}}, \bibinfo {author}
  {\bibfnamefont {E.}~\bibnamefont {Altman}}, \bibinfo {author} {\bibfnamefont
  {U.}~\bibnamefont {Schneider}}, \ and\ \bibinfo {author} {\bibfnamefont
  {I.}~\bibnamefont {Bloch}},\ }\href@noop {} {\bibfield  {journal} {\bibinfo
  {journal} {Science}\ }\textbf {\bibinfo {volume} {349}},\ \bibinfo {pages}
  {842} (\bibinfo {year} {2015})}\BibitemShut {NoStop}%
\bibitem [{\citenamefont {Bordia}\ \emph {et~al.}()\citenamefont {Bordia},
  \citenamefont {Lueschen}, \citenamefont {Hodgman}, \citenamefont {Schreiber},
  \citenamefont {Bloch},\ and\ \citenamefont {Schneider}}]{Bordia2015}%
  \BibitemOpen
  \bibfield  {author} {\bibinfo {author} {\bibfnamefont {P.}~\bibnamefont
  {Bordia}}, \bibinfo {author} {\bibfnamefont {H.~P.}\ \bibnamefont
  {Lueschen}}, \bibinfo {author} {\bibfnamefont {S.~S.}\ \bibnamefont
  {Hodgman}}, \bibinfo {author} {\bibfnamefont {M.}~\bibnamefont {Schreiber}},
  \bibinfo {author} {\bibfnamefont {I.}~\bibnamefont {Bloch}}, \ and\ \bibinfo
  {author} {\bibfnamefont {U.}~\bibnamefont {Schneider}},\ }\href@noop {}
  {\bibinfo  {journal} {arXiv:1509.00478}\ }\BibitemShut {NoStop}%
\bibitem [{\citenamefont {Smith}\ \emph {et~al.}()\citenamefont {Smith},
  \citenamefont {Lee}, \citenamefont {Richerme}, \citenamefont {Neyenhuis},
  \citenamefont {Hess}, \citenamefont {Hauke}, \citenamefont {Heyl},
  \citenamefont {Huse},\ and\ \citenamefont {Monroe}}]{Smith2015}%
  \BibitemOpen
\bibfield  {journal} {  }\bibfield  {author} {\bibinfo {author} {\bibfnamefont
  {J.}~\bibnamefont {Smith}}, \bibinfo {author} {\bibfnamefont
  {A.}~\bibnamefont {Lee}}, \bibinfo {author} {\bibfnamefont {P.}~\bibnamefont
  {Richerme}}, \bibinfo {author} {\bibfnamefont {B.}~\bibnamefont {Neyenhuis}},
  \bibinfo {author} {\bibfnamefont {P.~W.}\ \bibnamefont {Hess}}, \bibinfo
  {author} {\bibfnamefont {P.}~\bibnamefont {Hauke}}, \bibinfo {author}
  {\bibfnamefont {M.}~\bibnamefont {Heyl}}, \bibinfo {author} {\bibfnamefont
  {D.~A.}\ \bibnamefont {Huse}}, \ and\ \bibinfo {author} {\bibfnamefont
  {C.}~\bibnamefont {Monroe}},\ }\href@noop {} {\bibinfo  {journal}
  {arXiv:1508.07026}\ }\BibitemShut {NoStop}%
\bibitem [{\citenamefont {Polkovnikov}(2011)}]{Polkovnikov2011}%
  \BibitemOpen
\bibfield  {journal} {  }\bibfield  {author} {\bibinfo {author} {\bibfnamefont
  {A.}~\bibnamefont {Polkovnikov}},\ }\href@noop {} {\bibfield  {journal}
  {\bibinfo  {journal} {Ann. of Phys.}\ }\textbf {\bibinfo {volume} {326}},\
  \bibinfo {pages} {486} (\bibinfo {year} {2011})}\BibitemShut {NoStop}%
\bibitem [{\citenamefont {Luitz}\ \emph {et~al.}(2015)\citenamefont {Luitz},
  \citenamefont {Laflorencie},\ and\ \citenamefont {Alet}}]{Luitz2015}%
  \BibitemOpen
  \bibfield  {author} {\bibinfo {author} {\bibfnamefont {D.~J.}\ \bibnamefont
  {Luitz}}, \bibinfo {author} {\bibfnamefont {N.}~\bibnamefont {Laflorencie}},
  \ and\ \bibinfo {author} {\bibfnamefont {F.}~\bibnamefont {Alet}},\
  }\href@noop {} {\bibfield  {journal} {\bibinfo  {journal} {Phys. Rev. B}\
  }\textbf {\bibinfo {volume} {91}},\ \bibinfo {pages} {081103} (\bibinfo
  {year} {2015})}\BibitemShut {NoStop}%
\bibitem [{\citenamefont {Serbyn}\ \emph {et~al.}()\citenamefont {Serbyn},
  \citenamefont {Papic},\ and\ \citenamefont {Abanin}}]{Serbyn2015b}%
  \BibitemOpen
  \bibfield  {author} {\bibinfo {author} {\bibfnamefont {M.}~\bibnamefont
  {Serbyn}}, \bibinfo {author} {\bibfnamefont {Z.}~\bibnamefont {Papic}}, \
  and\ \bibinfo {author} {\bibfnamefont {D.~A.}\ \bibnamefont {Abanin}},\
  }\href@noop {} {\bibinfo  {journal} {arXiv:1507.01635}\ }\BibitemShut
  {NoStop}%
\bibitem [{\citenamefont {de~Roeck}\ \emph {et~al.}()\citenamefont {de~Roeck},
  \citenamefont {Huveneers}, \citenamefont {M{\"u}ller},\ and\ \citenamefont
  {Schiulaz}}]{Roeck2015}%
  \BibitemOpen
\bibfield  {journal} {  }\bibfield  {author} {\bibinfo {author} {\bibfnamefont
  {W.}~\bibnamefont {de~Roeck}}, \bibinfo {author} {\bibfnamefont
  {F.}~\bibnamefont {Huveneers}}, \bibinfo {author} {\bibfnamefont
  {M.}~\bibnamefont {M{\"u}ller}}, \ and\ \bibinfo {author} {\bibfnamefont
  {M.}~\bibnamefont {Schiulaz}},\ }\href@noop {} {\bibinfo  {journal}
  {arXiv:1506.01505}\ }\BibitemShut {NoStop}%
\bibitem [{\citenamefont {Lindblad}(1976)}]{Lindblad1976}%
  \BibitemOpen
\bibfield  {journal} {  }\bibfield  {author} {\bibinfo {author} {\bibfnamefont
  {G.}~\bibnamefont {Lindblad}},\ }\href@noop {} {\bibfield  {journal}
  {\bibinfo  {journal} {Comm. Math. Phys}\ }\textbf {\bibinfo {volume} {48}},\
  \bibinfo {pages} {119} (\bibinfo {year} {1976})}\BibitemShut {NoStop}%
\bibitem [{\citenamefont {Gardiner}\ and\ \citenamefont
  {Zoller}(2004)}]{Gardiner2004}%
  \BibitemOpen
  \bibfield  {author} {\bibinfo {author} {\bibfnamefont {C.}~\bibnamefont
  {Gardiner}}\ and\ \bibinfo {author} {\bibfnamefont {P.}~\bibnamefont
  {Zoller}},\ }\href@noop {} {\emph {\bibinfo {title} {{Quantum noise}}}}\
  (\bibinfo  {publisher} {Springer},\ \bibinfo {year} {2004})\BibitemShut
  {NoStop}%
\bibitem [{\citenamefont {Mott}(1969)}]{Mott1969}%
  \BibitemOpen
  \bibfield  {author} {\bibinfo {author} {\bibfnamefont {N.~F.}\ \bibnamefont
  {Mott}},\ }\href@noop {} {\bibfield  {journal} {\bibinfo  {journal} {Phil.
  Mag.}\ }\textbf {\bibinfo {volume} {19}},\ \bibinfo {pages} {835} (\bibinfo
  {year} {1969})}\BibitemShut {NoStop}%
\bibitem [{\citenamefont {Carmele}\ \emph {et~al.}(2015)\citenamefont
  {Carmele}, \citenamefont {Heyl}, \citenamefont {Kraus},\ and\ \citenamefont
  {Dalmonte}}]{Carmele2015}%
  \BibitemOpen
  \bibfield  {author} {\bibinfo {author} {\bibfnamefont {A.}~\bibnamefont
  {Carmele}}, \bibinfo {author} {\bibfnamefont {M.}~\bibnamefont {Heyl}},
  \bibinfo {author} {\bibfnamefont {C.}~\bibnamefont {Kraus}}, \ and\ \bibinfo
  {author} {\bibfnamefont {M.}~\bibnamefont {Dalmonte}},\ }\href@noop {}
  {\bibfield  {journal} {\bibinfo  {journal} {Phys. Rev. B}\ }\textbf {\bibinfo
  {volume} {92}},\ \bibinfo {pages} {195107} (\bibinfo {year}
  {2015})}\BibitemShut {NoStop}%
\bibitem [{\citenamefont {Genway}\ \emph {et~al.}(2014)\citenamefont {Genway},
  \citenamefont {Lesanovsky},\ and\ \citenamefont {Garrahan}}]{Genway2014}%
  \BibitemOpen
  \bibfield  {author} {\bibinfo {author} {\bibfnamefont {S.}~\bibnamefont
  {Genway}}, \bibinfo {author} {\bibfnamefont {I.}~\bibnamefont {Lesanovsky}},
  \ and\ \bibinfo {author} {\bibfnamefont {J.~P.}\ \bibnamefont {Garrahan}},\
  }\href@noop {} {\bibfield  {journal} {\bibinfo  {journal} {Phys. Rev. E}\
  }\textbf {\bibinfo {volume} {89}},\ \bibinfo {pages} {042129} (\bibinfo
  {year} {2014})}\BibitemShut {NoStop}%
\bibitem [{\citenamefont {Nandkishore}\ \emph {et~al.}(2014)\citenamefont
  {Nandkishore}, \citenamefont {Gopalakrishnan},\ and\ \citenamefont
  {Huse}}]{Nandkishore2014}%
  \BibitemOpen
  \bibfield  {author} {\bibinfo {author} {\bibfnamefont {R.}~\bibnamefont
  {Nandkishore}}, \bibinfo {author} {\bibfnamefont {S.}~\bibnamefont
  {Gopalakrishnan}}, \ and\ \bibinfo {author} {\bibfnamefont {D.~A.}\
  \bibnamefont {Huse}},\ }\href {\doibase 10.1103/PhysRevB.90.064203}
  {\bibfield  {journal} {\bibinfo  {journal} {Phys. Rev. B}\ }\textbf {\bibinfo
  {volume} {90}},\ \bibinfo {pages} {064203} (\bibinfo {year}
  {2014})}\BibitemShut {NoStop}%
\bibitem [{\citenamefont {Huse}\ \emph {et~al.}(2015)\citenamefont {Huse},
  \citenamefont {Nandkishore}, \citenamefont {Pietracaprina}, \citenamefont
  {Ros},\ and\ \citenamefont {Scardicchio}}]{Huse2015}%
  \BibitemOpen
  \bibfield  {author} {\bibinfo {author} {\bibfnamefont {D.~A.}\ \bibnamefont
  {Huse}}, \bibinfo {author} {\bibfnamefont {R.}~\bibnamefont {Nandkishore}},
  \bibinfo {author} {\bibfnamefont {F.}~\bibnamefont {Pietracaprina}}, \bibinfo
  {author} {\bibfnamefont {V.}~\bibnamefont {Ros}}, \ and\ \bibinfo {author}
  {\bibfnamefont {A.}~\bibnamefont {Scardicchio}},\ }\href@noop {} {\bibfield
  {journal} {\bibinfo  {journal} {Phys. Rev. B}\ }\textbf {\bibinfo {volume}
  {92}},\ \bibinfo {pages} {014203} (\bibinfo {year} {2015})}\BibitemShut
  {NoStop}%
\bibitem [{\citenamefont {Nandkishore}()}]{Nandkishore2015a}%
  \BibitemOpen
  \bibfield  {author} {\bibinfo {author} {\bibfnamefont {R.}~\bibnamefont
  {Nandkishore}},\ }\href@noop {} {\bibinfo  {journal} {arXiv:1506.05468}\
  }\BibitemShut {NoStop}%
\bibitem [{\citenamefont {Johri}\ \emph {et~al.}(2015)\citenamefont {Johri},
  \citenamefont {Nandkishore},\ and\ \citenamefont {Bhatt}}]{Johri2015}%
  \BibitemOpen
\bibfield  {journal} {  }\bibfield  {author} {\bibinfo {author} {\bibfnamefont
  {S.}~\bibnamefont {Johri}}, \bibinfo {author} {\bibfnamefont
  {R.}~\bibnamefont {Nandkishore}}, \ and\ \bibinfo {author} {\bibfnamefont
  {R.~N.}\ \bibnamefont {Bhatt}},\ }\href@noop {} {\bibfield  {journal}
  {\bibinfo  {journal} {Phys. Rev. Lett.}\ }\textbf {\bibinfo {volume} {114}},\
  \bibinfo {pages} {117401} (\bibinfo {year} {2015})}\BibitemShut {NoStop}%
\bibitem [{\citenamefont {\v{Z}nidari\v{c}}(2010)}]{Znidaric2010}%
  \BibitemOpen
  \bibfield  {author} {\bibinfo {author} {\bibfnamefont {M.}~\bibnamefont
  {\v{Z}nidari\v{c}}},\ }\href@noop {} {\bibfield  {journal} {\bibinfo
  {journal} {New J. Phys.}\ }\textbf {\bibinfo {volume} {12}},\ \bibinfo
  {pages} {043001} (\bibinfo {year} {2010})}\BibitemShut {NoStop}%
\bibitem [{\citenamefont {Johansson}\ \emph {et~al.}(2012)\citenamefont
  {Johansson}, \citenamefont {Nation},\ and\ \citenamefont {Nori}}]{Qutip1}%
  \BibitemOpen
  \bibfield  {author} {\bibinfo {author} {\bibfnamefont {J.}~\bibnamefont
  {Johansson}}, \bibinfo {author} {\bibfnamefont {P.}~\bibnamefont {Nation}}, \
  and\ \bibinfo {author} {\bibfnamefont {F.}~\bibnamefont {Nori}},\ }\href@noop
  {} {\bibfield  {journal} {\bibinfo  {journal} {Comp. Phys. Comm.}\ }\textbf
  {\bibinfo {volume} {183}},\ \bibinfo {pages} {1760} (\bibinfo {year}
  {2012})}\BibitemShut {NoStop}%
\bibitem [{\citenamefont {Johansson}\ \emph {et~al.}(2013)\citenamefont
  {Johansson}, \citenamefont {Nation},\ and\ \citenamefont {Nori}}]{Qutip2}%
  \BibitemOpen
  \bibfield  {author} {\bibinfo {author} {\bibfnamefont {J.}~\bibnamefont
  {Johansson}}, \bibinfo {author} {\bibfnamefont {P.}~\bibnamefont {Nation}}, \
  and\ \bibinfo {author} {\bibfnamefont {F.}~\bibnamefont {Nori}},\ }\href@noop
  {} {\bibfield  {journal} {\bibinfo  {journal} {Comp. Phys. Comm.}\ }\textbf
  {\bibinfo {volume} {184}},\ \bibinfo {pages} {1234} (\bibinfo {year}
  {2013})}\BibitemShut {NoStop}%
\bibitem [{\citenamefont {Hauke}\ and\ \citenamefont {Heyl}(2015)}]{Hauke2014}%
  \BibitemOpen
  \bibfield  {author} {\bibinfo {author} {\bibfnamefont {P.}~\bibnamefont
  {Hauke}}\ and\ \bibinfo {author} {\bibfnamefont {M.}~\bibnamefont {Heyl}},\
  }\href@noop {} {\bibfield  {journal} {\bibinfo  {journal} {Phys. Rev. B}\
  }\textbf {\bibinfo {volume} {92}},\ \bibinfo {pages} {134204} (\bibinfo
  {year} {2015})}\BibitemShut {NoStop}%
\bibitem [{\citenamefont {Dutta}\ \emph {et~al.}(2015)\citenamefont {Dutta},
  \citenamefont {Gajda}, \citenamefont {Hauke}, \citenamefont {Lewenstein},
  \citenamefont {L{\"u}hmann}, \citenamefont {Malomed}, \citenamefont
  {Sowi{\'n}ski},\ and\ \citenamefont {Zakrzewski}}]{Dutta2015}%
  \BibitemOpen
  \bibfield  {author} {\bibinfo {author} {\bibfnamefont {O.}~\bibnamefont
  {Dutta}}, \bibinfo {author} {\bibfnamefont {M.}~\bibnamefont {Gajda}},
  \bibinfo {author} {\bibfnamefont {P.}~\bibnamefont {Hauke}}, \bibinfo
  {author} {\bibfnamefont {M.}~\bibnamefont {Lewenstein}}, \bibinfo {author}
  {\bibfnamefont {D.-S.}\ \bibnamefont {L{\"u}hmann}}, \bibinfo {author}
  {\bibfnamefont {B.~A.}\ \bibnamefont {Malomed}}, \bibinfo {author}
  {\bibfnamefont {T.}~\bibnamefont {Sowi{\'n}ski}}, \ and\ \bibinfo {author}
  {\bibfnamefont {J.}~\bibnamefont {Zakrzewski}},\ }\href@noop {} {\bibfield
  {journal} {\bibinfo  {journal} {Rep. Prog. Phys.}\ }\textbf {\bibinfo
  {volume} {78}},\ \bibinfo {pages} {066001} (\bibinfo {year}
  {2015})}\BibitemShut {NoStop}%
\bibitem [{\citenamefont {Sarkar}\ \emph {et~al.}(2014)\citenamefont {Sarkar},
  \citenamefont {Langer}, \citenamefont {Schachenmayer},\ and\ \citenamefont
  {Daley}}]{Sarkar2014}%
  \BibitemOpen
  \bibfield  {author} {\bibinfo {author} {\bibfnamefont {S.}~\bibnamefont
  {Sarkar}}, \bibinfo {author} {\bibfnamefont {S.}~\bibnamefont {Langer}},
  \bibinfo {author} {\bibfnamefont {J.}~\bibnamefont {Schachenmayer}}, \ and\
  \bibinfo {author} {\bibfnamefont {A.~J.}\ \bibnamefont {Daley}},\ }\href
  {\doibase 10.1103/PhysRevA.90.023618} {\bibfield  {journal} {\bibinfo
  {journal} {Phys. Rev. A}\ }\textbf {\bibinfo {volume} {90}},\ \bibinfo
  {pages} {023618} (\bibinfo {year} {2014})}\BibitemShut {NoStop}%
\bibitem [{\citenamefont {M{\o}lmer}\ \emph {et~al.}(1993)\citenamefont
  {M{\o}lmer}, \citenamefont {Castin},\ and\ \citenamefont
  {Dalibard}}]{Qjumps1}%
  \BibitemOpen
  \bibfield  {author} {\bibinfo {author} {\bibfnamefont {K.}~\bibnamefont
  {M{\o}lmer}}, \bibinfo {author} {\bibfnamefont {Y.}~\bibnamefont {Castin}}, \
  and\ \bibinfo {author} {\bibfnamefont {J.}~\bibnamefont {Dalibard}},\ }\href
  {\doibase 10.1364/JOSAB.10.000524} {\bibfield  {journal} {\bibinfo  {journal}
  {J. Opt. Soc. Am. B}\ }\textbf {\bibinfo {volume} {10}},\ \bibinfo {pages}
  {524} (\bibinfo {year} {1993})}\BibitemShut {NoStop}%
\bibitem [{\citenamefont {Plenio}\ and\ \citenamefont
  {Knight}(1998)}]{Qjumps2}%
  \BibitemOpen
  \bibfield  {author} {\bibinfo {author} {\bibfnamefont {M.~B.}\ \bibnamefont
  {Plenio}}\ and\ \bibinfo {author} {\bibfnamefont {P.~L.}\ \bibnamefont
  {Knight}},\ }\href {\doibase 10.1103/RevModPhys.70.101} {\bibfield  {journal}
  {\bibinfo  {journal} {Rev. Mod. Phys.}\ }\textbf {\bibinfo {volume} {70}},\
  \bibinfo {pages} {101} (\bibinfo {year} {1998})}\BibitemShut {NoStop}%
\bibitem [{\citenamefont {Daley}(2014)}]{Daley2014}%
  \BibitemOpen
  \bibfield  {author} {\bibinfo {author} {\bibfnamefont {A.~J.}\ \bibnamefont
  {Daley}},\ }\href@noop {} {\bibfield  {journal} {\bibinfo  {journal} {Adv.
  Phys.}\ }\textbf {\bibinfo {volume} {63}},\ \bibinfo {pages} {77} (\bibinfo
  {year} {2014})}\BibitemShut {NoStop}%
\bibitem [{\citenamefont {Garrahan}\ and\ \citenamefont
  {Chandler}(2002)}]{Garrahan2002}%
  \BibitemOpen
  \bibfield  {author} {\bibinfo {author} {\bibfnamefont {J.~P.}\ \bibnamefont
  {Garrahan}}\ and\ \bibinfo {author} {\bibfnamefont {D.}~\bibnamefont
  {Chandler}},\ }\href@noop {} {\bibfield  {journal} {\bibinfo  {journal}
  {Phys. Rev. Lett.}\ }\textbf {\bibinfo {volume} {89}},\ \bibinfo {pages}
  {035704} (\bibinfo {year} {2002})}\BibitemShut {NoStop}%
\end{thebibliography}
%

\end{document}